% =======================LETRAS HUECAS============================
\newfam\msbfam
\font\twlmsb=msbm10 at 12pt
\font\eightmsb=msbm10 at 8pt
\font\sixmsb=msbm10 at 6pt
\textfont\msbfam=\twlmsb
\scriptfont\msbfam=\eightmsb
\scriptscriptfont\msbfam=\sixmsb
\def\cj{\fam\msbfam}

\def\C{{\cj C}}

\def\h{{\cj H}}

\def\R{{\cj R}}

\def\Z{{\cj Z}}

\centerline{\bf THE CPT GROUP OF THE DIRAC FIELD}

\

\centerline{\bf M. Socolovsky} 

\
 
\centerline{\it Instituto de Ciencias Nucleares, Universidad Nacional Aut\'onoma de M\'exico} 
\centerline{\it Circuito Exterior, Ciudad Universitaria, 04510, M\'exico D.F., M\'exico} 

\

\

{\it Using the standard representation of the Dirac equation, we show that, up to signs, there exist only {\bf two sets} of consistent solutions for the matrices of charge conjugation ($C$), parity ($P$), and time reversal ($T$), which give the transformation of fields $\psi_C(x)=C\bar{\psi}^T(x)$, $\psi_\Pi(x_\Pi)=P\psi(x)$ and $\psi_\tau (x)=T\psi(x_\tau)^*$, where $x_\Pi=(t,-\vec{x})$ and $
x_\tau=(-t,\vec{x})$. These sets are given by $C=\pm\gamma^2\gamma_0$, $P=\pm i\gamma_0$, $T=\pm i\gamma^3\gamma^1$ and $C=\pm i\gamma^2\gamma_0$, $P=\pm i\gamma_0$, $T=\pm \gamma^3\gamma^1$. Then $P^2=-1$, and two succesive applications of the parity transformation to fermion fields {\bf necessarily} amounts to a $2\pi$ rotation. Each of these sets generates a non abelian group of sixteen elements, respectively $G_\theta^{(1)}$ and $G_\theta^{(2)}$, which are non isomorphic subgroups of the Dirac algebra, which, being a Clifford algebra, gives a geometric nature to the generators, in particular to charge conjugation. It turns out that $G_\theta^{(1)}\cong DH_8 \times \Z_2 \subset S_6$ and $G_\theta ^{(2)} \cong 16E \subset S_8$, where $DH_8$ is the dihedral group of eight elements, the group of symmetries of the square, and $16E$ is a non trivial extension of $DH_8$ by $\Z_2$, isomorphic to a semidirect product of these groups; $S_6$ and $S_8$ are the symmetric groups of six and eight elements. The matrices are also given in the Weyl representation, suitable for taking the massless limit, and in the Majorana representation, describing self-conjugate fields. Instead, the quantum operators $\bf C$, $\bf P$ and $\bf T$, acting on the Hilbert space, generate a unique group $G_\Theta$, which we call {\bf the} $\bf CPT$ {\bf group of the Dirac field}. This group, however, is compatible only with the second of the above two matrix solutions, namely with $G_\theta^{(2)}$, which  is then called {\bf the matrix} $CPT$ {\bf group}. It turns out that $G_\Theta \cong DC_8\times \Z_2\subset S_{10}$, where $DC_8$ is the dicyclic group of eight elements and $S_{10}$ is the symmetric group of ten elements. Since $DC_8\cong Q$, the quaternion group, and $\Z_2 \cong S^0$, the 0-sphere, then $G_\Theta \cong Q\times S^0$.}

\

\

{\bf Key words}: discrete symmetries; Dirac equation; quantum field theory; finite groups.

\

\

{\bf 1. Introduction}

\

Let $(a,\omega)$ be an element of the Poincar\'e group ${\cal P}$, the semidirect sum of ${\cal T}$, the translations, and ${\cal L}$, the Lorentz group, of the 4-dimensional Minkowski space-time. If $u(x)$ is a linear field operator on the Hilbert space ${\cal H}$, then under $(a,\omega)$, $u(x)$ transforms as $$u^\prime(x^\prime)=\Lambda (\omega)u(x) \eqno{(1)}$$ where $\Lambda(\omega)$ is the $n\times n$ matrix representation of $(a, \omega)$ acting on the $n$ components of $u(x)$, and $x^\prime=(a,\omega)\cdot x=\omega x+a$. The state vector $\Psi \in {\cal H}$ of the system of fields, on the other hand, transforms as $$\Psi^\prime=U(a,\omega)\Psi \eqno{(2)}$$ where $U(a,\omega)$ is the operator representing $(a,\omega)$ in the Hilbert space. So, the mean value of $u(x)$ in the state $\Psi^\prime$ is given by $$(\Psi^\prime,u(x)\Psi^\prime)=(U(a,\omega)\Psi,u(x)U(a,\omega)\Psi)=(\Psi,U^{\dag} (a,\omega)u(x)U(a,\omega)\Psi)=(\Psi,u^\prime(x)\Psi) \eqno{(3a)}$$ with $$u^\prime(x)=U^{\dag}(a,\omega)u(x)U(a,\omega), \ U^{\dag}(a,\omega)=U^{-1}(a,\omega) \eqno {(4a)}$$ for {\it unitary} $U$, and $$(\Psi^\prime,u(x)\Psi^\prime)=(u^{\dag}(x)V(a,\omega)\Psi,V(a,\omega)\Psi)=(V(a,\omega)V^{\dag}(a,\omega)u^{\dag}(x)V(a,\omega)\Psi,V(a,\omega)\Psi)$$  $$=(\Psi,V^{\dag}(a,\omega)u^{\dag}(x)V(a,\omega)\Psi)=(\Psi,(V^{\dag}(a,\omega)u(x)V(a,\omega))^{\dag}\Psi)=(\Psi,u^\prime(x)\Psi) \eqno{(3b)}$$ with $$u^\prime(x)=V^{\dag}(a,\omega)u^{\dag}(x)V(a,\omega), \ V^{\dag}(a,\omega)=V^{-1}(a,\omega) \eqno{(4b)}$$ for {\it antiunitary} $U\equiv V$.

The left hand sides of both (3a) and (3b) are the analogues of the expectation value of a time independent operator in the Schr\"odinger picture of non relativistic quantum mechanics, while the respective right hand sides correspond to the Heisenberg picture. Comparing (1) with (4a) and (4b) one obtains the compatibility conditions (Bogoliubov and Shirkov, 1980) $$u^\prime(x)=\Lambda(\omega)u((a,\omega)^{-1}\cdot x)=U^{\dag}(a,\omega)u(x)U(a,\omega) \eqno{(5a)}$$ for unitary $U$, and $$u^\prime(x)=\Lambda(\omega)u((a,\omega)^{-1}\cdot x)=(V^{\dag}(a,\omega)u(x)V(a,\omega))^{\dag} \eqno{(5b)}$$ for antiunitary $V$. Through the matrices $\Lambda(\omega)$, (5a) and (5b) define the action of the operators $U$ and $V$ on the quantum field operators $u(x)$.

Corresponding to the transformations of parity ($(a,\omega)=(0,\Pi)$) and time reversal ($(a,\omega)=(0,\tau)$) we have the operators $U(0,\Pi)={\bf P}$ (unitary) and $V(0,\tau)={\bf T}$ (antiunitary), and the matrices $\Lambda(\Pi)=P$ and $\Lambda(\tau)=T$. To charge conjugation ($c$), which corresponds to particle-antiparticle interchange, and which is not a space-time transformation {\it i.e.} $c \notin {\cal P}$, it corresponds the unitary operator $U={\bf C}$ and the matrix $\Lambda=C$.

In the following we shall restrict the discussion to the Dirac field describing massive spin ${1}\over {2}$ particles; so $u(x)=\psi(x)$ with $n=4$. However, as shown in section 8, the results are independent of the value of the mass, and in particular they hold in the massless limit. The Dirac algebra $D^{16}$, isomorphic to $\C(4)$, the algebra of 4$\times$4 matrices with complex entries, is a complex Clifford algebra with canonical basis given by $$\{1, \gamma_0, \ \gamma^1, \ \gamma^2, \ \gamma^3, \ \gamma_0 \ \gamma^1, \ \gamma_0 \gamma^2, \ \gamma_0 \gamma^3, \ \gamma^1 \gamma^2, \ \gamma^2 \gamma^3, \ \gamma^3 \gamma^1, \ \gamma_0 \gamma^1 \gamma^2, \ \gamma^1 \gamma^2 \gamma^3, \ \gamma^2 \gamma^3 \gamma_0, \ \gamma^3 \gamma_0 \gamma^1, \ \gamma_0 \gamma^1 \gamma^2 \gamma^3 \};$$ $D^{16}$ is the complexification of both $\h(2)=\{2\times 2 \ quaternionic \ matrices\}$ which is the real Clifford algebra of $\R^4$ with metric $\eta_{\mu\nu}=diag(1,-1,-1,-1)$, and of $\R(4)=\{4\times 4 \ real \ matrices\}$ which is the real Clifford algebra of $\R^4$ with metric $\tilde{\eta}_{\mu\nu}=(-1,1,1,1)$; as real algebras, $\h(2)$ is not isomorphic to $\R(4)$ though the corresponding metrics are physically equivalent (Socolovsky, 2001). This implies the need of complexification to define the physical Dirac algebra. 

For completeness, in the final section, we shall give the $C$, $P$, $T$ and $CPT$ matrices in the Weyl representation, which is adequate for the massless limit, and in the Majorana representation, for self-conjugate fermions. Obviously, the group structures of section 7 are independent of the representation used. 

\

\

{\bf 2. Parity}

\

Starting from the Dirac equation $$(i\gamma^{\mu}\partial_{\mu}-m)\psi(x)=0$$ and making the transformation $\Pi:x^{\mu}=(t,\vec{x})\to x^\mu_\Pi=(t,-\vec{x})=\omega^{\mu}_{\nu}x^{\nu}$ with $\omega^{\mu}_{\nu}=diag (1,-1,-1,-1)$, one looks for the 4$\times$4 invertible matrix $P$ such that $\psi_\Pi (x_\Pi) =P\psi(x)$ satisfies the equation $$(i\gamma^{\mu}\partial^\prime_{\mu}-m)\psi_\Pi(x_\Pi)=0$$ {\it i.e.} $$(i(\gamma_0\partial_0-\gamma^{i}\partial _i)-m)P\psi(x)=0$$ with $\partial _\mu={{\partial}\over {\partial {x^{\mu}}}}$. Multiplying from the left by $P^{-1}$ one obtains $$(i(P^{-1}\gamma_0P\partial_0-P^{-1}\gamma^{i}P) \partial_i-m)\psi(x)=0$$ which implies the constraints on $P$: $$P^{-1}\gamma_0 P=\gamma_0 \ \ and \ \ P^{-1}\gamma^kP=-\gamma^k$$ {\it i.e.} $$P^{-1}\gamma^\mu P=\omega^\mu _\nu \gamma^\nu \ \ or \ \ [P,\gamma_0]=\{P,\gamma^k\}=0, \ k=1,2,3.\eqno {(6)}$$ 

Using the standard or Dirac-Pauli (DP) form of the $\gamma$-matrices, namely, $\gamma_0=\pmatrix{1 & 0 \cr 0 & -1 \cr}$ and $\gamma^k=\pmatrix{0 & \sigma_k \cr -\sigma_k & 0 \cr}$, with $\sigma_1=\pmatrix{0 & 1 \cr 1 & 0 \cr}$, $\sigma_2=\pmatrix{0 & -i \cr i & 0 \cr}$, and $\sigma_3=\pmatrix{1 & 0 \cr 0 & -1 \cr}$ the Pauli matrices, one easily verifies that the unique solution to (6) is $$P=z\gamma_0, \ z\in \C^*=\C-\{0\}, \eqno{(7)}$$ with $P\in D^{16}$. (See Appendix 1.)

\

Since $det(\Pi ^2)=1$, then $\Pi ^2$ must be a rotation. Since under a rotation of 360$^{\circ}$ spinors change sign, we have two possibilities for $P^2=z^2\gamma_0^2=z^21$ (Racah, 1937; Yang and Tiomno, 1950; Wick {\it et al}, 1952; Berestetskii {\it et al}, 1982; Sternberg, 1994):$$P^2=+1 \ \Rightarrow \ z=\pm 1 \eqno{(8a)}$$ or $$P^2=-1 \ \Rightarrow \ z=\pm i. \eqno{(8b)}$$ In the first case, $P^2$ is equivalent to a $0^{\circ}$ rotation {\it i.e.} to no rotation at all, and one has: $$P=\pm \gamma_0, \ P^{\dag}=P=P^{-1}=P^\sim =P^*, \ det(P)=1, \ tr(P)=0. \eqno{(9a)}$$ ($\sim$ denotes the transpose matrix.) In the second case, $P$ amounts to a $2\pi$ rotation:$$P=\pm i\gamma_0, \ P^{\dag}=-P=P^{-1}=-P^\sim =P^*, \ det(P)=1, \ tr(P)=0. \eqno {(9b)}$$ 

\

For later use, we find the parity transformation for the Dirac conjugate spinor: from $\psi_\Pi(x_\Pi)=P\psi(x)$ one has $\psi_\Pi(x_\Pi)^{\dag}\gamma_0=\psi(x)^{\dag}P^{\dag}\gamma_0=\psi(x)^{\dag}\gamma_0 P^{-1}=\bar{\psi}(x)P^{-1}$ and the l.h.s. defines $\bar{\psi}_\Pi(x_\Pi)$ {\it i.e.} $$\bar{\psi}_\Pi(x_\Pi)=\bar{\psi}(x)P^{-1}. \eqno (10)$$

\

Formula (5a) in this case is $$\psi_\Pi(t,\vec{x})=P\psi(t,-\vec{x})={\bf P}^{\dag}\psi(t,\vec{x}){\bf P}, \eqno{(11)}$$ and therefore $${\bf P}^{\dag}({\bf P}^{\dag}\psi(t,\vec{x}){\bf P}){\bf P}={\bf P}^{\dag 2}\psi(t,\vec{x}){\bf P}^2={\bf P}^{\dag}(P\psi(t,-\vec{x})){\bf P}=P(P\psi(t,\vec{x}))=P^2\psi(t,\vec{x}) $$ {\it i.e.} $${\bf P}^{\dag 2}\psi(t,\vec{x}){\bf P}^2=\pm \psi(t,\vec{x}), \eqno{(12)}$$ corresponding to the two solutions (8a) and (8b), which imply $$\psi_{\Pi^2}=\pm \psi.$$

\

{\bf 3. Charge conjugation}

\

The Dirac equation for an electric charge $q=-\vert e \vert$ in an electromagnetic potential $A_\mu$ is given by $$(i\gamma^\mu \partial_\mu+\vert e \vert \gamma^\mu A_\mu-m)\psi(x)=0. \eqno{(13)}$$ Taking the complex conjugate of this equation, multiplying from the left by $C\gamma_0$ where $C$ is a matrix in $GL_4(\C)$, and inserting the unit matrix, one obtains $$((i\partial_\mu-\vert e \vert A_\mu)(C\gamma_0)\gamma^{\mu*}(C\gamma_0)^{-1}+m)(C\gamma_0)\psi^*=0.$$ Defining the {\it charge conjugate spinor} $$\psi_{C}=C\gamma_0\psi^* \eqno{(14)}$$ and imposing the constraint on $C$: $$(C\gamma_0)\gamma^{\mu*}(C\gamma_0)^{-1}=-\gamma^\mu, \eqno{(15)}$$ $\psi_{C}$ obeys the equation $$(i\gamma^\mu \partial_\mu- \vert e \vert \gamma^\mu A_\mu-m)\psi_{C}=0. \eqno{(16)}$$ Then $\psi_{C}$ describes particles with the same mass but with the opposite charge. Notice that if one completes the charge conjugation operation, namely one also changes $A_\mu \to -A_\mu$, then (16) becomes $$(i\gamma^\mu \partial _\mu + \vert e \vert \gamma^\mu A_\mu -m)\psi_{C}=0 \eqno{(13a)}$$ which exhibits the complete symmetry of quantum electrodynamics under the operation of charge conjugation. 

Since $\gamma_0\gamma^{\mu*}\gamma_0=\gamma^{\mu \sim}$, the constraint (15) is equivalent to $$C\gamma^{\mu \sim}C^{-1}=-\gamma^\mu. \eqno {(15a)}$$ Also, the Dirac conjugate spinor is $\bar{\psi}=\psi^{\dag}\gamma_0$ and therefore $\bar{\psi}^\sim=\gamma_0^\sim\psi ^{\dag \sim}=\gamma_0 \psi ^*$, so for the charge conjugate spinor one has $$\psi_C=C\bar{\psi}^\sim . \eqno{(14a)}$$ (15a) is equivalent to $$[C,\gamma^\mu]=0, \ \mu=1,3; \ \{C,\gamma^\mu\}=0, \ \mu=0,2 \eqno{(15b)}$$ which in turn implies $$[C,\gamma_5]=0, \ \gamma_5=-i\gamma_0\gamma^1 \gamma^2 \gamma^3=\pmatrix{0&-1 \cr -1&0 \cr}. \eqno{(17)}$$ 

It can be easily shown that the unique matrix $C$ which solves the constraints is of the form $$C=\eta \gamma^2 \gamma_0=\eta \pmatrix{0 & 0 & 0 & i \cr 0 & 0 & -i & 0 \cr 0 & i & 0 & 0 \cr -i & 0 & 0 & 0 \cr}\in D^{16} \eqno{(18)}$$ with $\eta  \in \C^*$; in particular $C^2=\eta^2 1, \ det(C)=\eta^4$ and $tr(C)=0$. (See Appendix 1.) 

\

A second application of the charge conjugation transformation leads to $$(\psi_C)_C=\psi_{C^2}=C\bar{\psi}_C^\sim=C(\psi_C^{\dag}\gamma_0)^\sim=C\gamma_0\psi_C^*=C\gamma_0C^*\gamma_0\psi=-CC^*\psi=-\vert \eta \vert^2\gamma^2\gamma_0\gamma^{2*}\gamma_0\psi$$ $$=-\vert \eta \vert ^2(\gamma^2)^2(\gamma_0)^2\psi=\vert \eta \vert ^2\psi$$ and therefore $$\psi_{C^2}=\psi \eqno{(19)}$$ since the effect on $\psi$ can be, at most, a multiplication by a phase. Then $\eta\in U(1)$ and $C$ is {\it unitary}: $CC^{\dag}=\eta \gamma^2\gamma_0\bar{\eta}\gamma_0 \gamma^{2{\dag}}=\vert \eta \vert^2\gamma^2 \gamma^{2{\dag}}=-\vert \eta \vert ^2(\gamma^2)^2=\vert \eta \vert^2 1=1$. On the other hand, for the transformation of the Dirac conjugate spinor one has $$\bar{\psi}_C=(\psi^{\dag}\gamma_0)_C=\psi^{\dag}_C\gamma_0=(C\gamma_0 \psi^*)^{\dag}\gamma_0=\psi^{*\dag}\gamma_0 C^{\dag}\gamma_0=\psi^\sim \gamma_0 \bar{\eta}(\gamma^2 \gamma_0)^{\dag}\gamma_0=-\bar{\eta}\psi^\sim \gamma^2 \gamma_0=-\bar{\eta}^2\psi^\sim\eta \gamma^2 \gamma_0$$ $$=-\bar{\eta}^2 \psi^\sim C \eqno{(14b)}$$ and then $$(\bar{\psi}\psi)_C=\bar{\psi}_C\psi_C=-\bar{\eta}^2\psi^\sim CC\bar{\psi}^\sim=-\bar{\eta}^2C^2(\bar{\psi}\psi)^\sim=-\bar{\eta}^2C^2\bar{\psi}\psi=-(\bar{\eta}\eta)^2\bar{\psi}\psi=-\vert \eta \vert ^4\bar{\psi}\psi=-\bar{\psi}\psi \eqno{(20)}$$ which is expected since $\bar{\psi}\psi$ is the charge density operator. Moreover, comparing (14b) with (14a), a symmetry consideration makes natural the assumption $$\bar{\psi}_C=\pm \psi^\sim C \eqno{(14c)}$$ {\it i.e.} $\eta^2=\pm 1$ which implies $\eta=\pm 1, \pm i$. Then $C$ becomes {\it unimodular}: $det(C)=1$, and one ends with the following two possibilities:
$$i) \ \eta=\pm 1 \Rightarrow C^2=1 \ and \ C=\pm\gamma^2 \gamma_0=C^{-1}=C^{\dag}=-C^\sim =-C^*, \eqno{(21a)}$$ $$\bar{\psi}_C=-\psi^\sim C; \eqno{(14d)}$$ $$ii) \ \eta=\pm i \Rightarrow C^2=-1 \ and \ C=\pm i \gamma^2 \gamma_0=-C^{-1}=-C^{\dag}=-C^\sim =C^*, \eqno{(21b)}$$ $$\bar{\psi}_C=\psi^\sim C. \eqno{(14e)}$$ 
We shall see that both i) and ii) for $C$ are consistent with $P^2=-1$ but not with $P^2=1$. This result, independently of the explicit form of $C$, was found for the first time by Racah (Racah, 1937); see also Schweber (Schweber, 1961), from whose analysis one can also obtain that $sign(C^2)=sign(T^2)$ (see below). The consistency of ii) for $C$ with $P^2=-1$ is explicitly shown by Capri (Capri, 2002). 

\

Formula (5a) in this case is $$\psi_C(x)=C\bar{\psi}^\sim(x)={\bf C}^{\dag}\psi(x){\bf C}, \eqno{(22)}$$ and therefore $${\bf C}^{\dag}({\bf C}^{\dag}\psi(x){\bf C}){\bf C}={\bf C}^{\dag 2}\psi(x){\bf C}^2={\bf C}^{\dag}(C\bar{\psi}^\sim(x)){\bf C}={\bf C}^{\dag}(C\gamma_0\psi^*(x)){\bf C}=(C\gamma_0\psi^*(x))_C$$  $$=C\gamma_0(C\gamma_0\psi^*(x))^*=C\gamma_0C^*\gamma_0\psi(x)=\gamma^2\gamma_0\gamma_0\gamma^{2*}\gamma_0\gamma_0\psi(x)=-(\gamma^2)^2\psi(x)$$ {\it i.e.} $${\bf C}^{\dag 2}\psi(x){\bf C}^2=\psi(x), \eqno{(23)}$$ as it must be, cf. (19).

\

\

{\bf 4. Fixing the square of} $P$

\

Let $(\psi_C)_\Pi(x_\Pi)$ be the parity transformed of the charge conjugate spinor; from (14a) and (10) one has $$(\psi_C)_\Pi(x_\Pi)=C\bar{\psi}_\Pi(x_\Pi)^\sim=C(\bar{\psi}(x)P^{-1})^\sim=C(P^{-1})^\sim\bar{\psi}(x)^\sim=C(P^{-1})^\sim C^{-1}C\bar{\psi}(x)^\sim $$ $=C(P^{-1})^\sim C^{-1}\psi_C(x)$, which must equal $P\psi_C(x)$; then $C$ and $P$ must satisfy $$C(P^{-1})^\sim C^{-1}=P. \eqno{(24)}$$

Consider the two possibilities for $C$ in section 3: 

\

i) $C(P^{-1})^\sim C^{-1}=\gamma^2\gamma_0(P^{-1})^\sim\gamma^2\gamma_0=\{\matrix{\gamma^2\gamma_0(\pm \gamma_0)\gamma^2\gamma_0=\mp \gamma_0=-P \cr \gamma^2\gamma_0(\mp i\gamma_0)\gamma^2\gamma_0=\pm i\gamma_0=P}$

which implies $P=\pm i\gamma_0$;

\

ii) $C(P^{-1})^\sim C^{-1}=(-i\gamma^2\gamma_0)(P^{-1})^\sim (i\gamma^2\gamma_0)=\{\matrix{\gamma^2\gamma_0(\pm\gamma_0)\gamma^2\gamma_0=\mp \gamma_0=-P \cr \gamma^2\gamma_0(\mp i\gamma_0)\gamma^2\gamma_0=\pm i\gamma_0=P}$

which also implies $P=\pm i\gamma_0$. 

\

So $$P \ and \ C \ are \ compatible \ if \ and \ only \ if \ P=\pm i \gamma_0, \ which \ implies \  P^2=-1.$$ 
Then $$\psi_{\Pi^2}=-\psi \ or, equivalently, \ {\bf P}^{{\dag}2}\psi(t,\vec{x}){\bf P}^2=-\psi(t,\vec{x}). \eqno{(12a)}$$

\

\

{\bf 5. Time reversal}

\

We start again from the free Dirac equation $$(i(\gamma_0 {{\partial}\over {\partial t}}+\gamma^k{{\partial}\over{\partial x^k}})-m)\psi(t,\vec{x})=0,$$ change $t\to -t$ and take the complex conjugate: $$(i(\gamma_0^*{{\partial}\over {\partial t}}-\gamma^{k*}{{\partial}\over{\partial x^k}})-m)\psi(-t,\vec{x})^*=0$$ with $\gamma_0^*=\gamma_0$, $\gamma^{k*}=\gamma^k$ for $k=1,3$ and $\gamma^{2*}=-\gamma^2$. Let $T$ be a 4$\times$4 invertible matrix in $\C(4)$ such that $$T\gamma_0T^{-1}=\gamma_0, \ T\gamma^{k*}T^{-1}=-\gamma^k. \eqno{(25)}$$ Then $$\psi_\tau(t,\vec{x})=T\psi(-t,\vec{x})^* \eqno{(26)}$$ obeys the Dirac equation $(i\gamma^\mu \partial _\mu-m)\psi_\tau(x)=0$. It is then easy to show that the solution of (25) is $$T=w \gamma^3\gamma^1=w \pmatrix{0 & -1 & 0 & 0 \cr 1 & 0 & 0 & 0 \cr 0 & 0 &0 & -1 \cr 0 & 0 & 1 & 0 \cr}, \ w\in \C^*. \eqno{(27)}$$ Clearly, $T\in D^{16}$. Also, $det(T)=w^4$ and $tr(T)=0$. (See Appendix 1.)

\

If we apply $\tau$ two times, we obtain $$\psi(t,\vec{x})\to \psi_\tau(t,\vec{x})=T\psi(-t,\vec{x})^*\to T(T\psi(t,\vec{x})^*)^*=TT^*\psi(t,\vec{x})$$ {\it i.e.} $\psi_{\tau^2}=TT^*\psi$. But $$TT^*=w\gamma^3\gamma^1 w^*\gamma^3\gamma^1=-\vert w \vert ^2(\gamma^3)^2(\gamma^1)^2=-\vert w \vert^2 1,$$ so $\psi_{\tau^2}=-\vert w \vert ^2\psi$ and therefore $$\psi_{\tau^2}=-\psi \eqno{(28)}$$ by a similar argument to the one used for $C$. So, $TT^*=-1$ {\it i.e.} $T^*=-T^{-1}$ and $w\in U(1): \ T=e^{i\lambda}\gamma^3\gamma^1$ and $T^{\dag}=-e^{-i\lambda}\gamma^3\gamma^1$. 

\

Formula (5b) in this case is $$\psi_\tau ^{\dag}(t,\vec{x})=\psi(-t,\vec{x})^{\dag}T^{\dag}=({\bf T}^{\dag}\psi(t,\vec{x}){\bf T})^{\dag}={\bf T}^{\dag}\psi(t,\vec{x})^{\dag}{\bf T}, \eqno{(29)}$$ 
which is equivalent to $$\psi_\tau(t,\vec{x})=T\psi(-t,\vec{x})={\bf T}^{\dag}\psi(t,\vec{x}){\bf T}; \eqno{(29a)}$$ therefore $${\bf T}^{\dag}({\bf T}^{\dag}\psi(t,\vec{x})^{\dag}{\bf T}){\bf T}={\bf T}^{\dag 2}\psi(t,\vec{x})^{\dag}{\bf T}^2={\bf T}^{\dag}\psi(-t,\vec{x})^{\dag}T^{\dag}{\bf T}={\bf T}^{\dag}\psi(-t,\vec{x})^{\dag}{\bf T}T^{{\dag}*}$$ $$=\psi(t,\vec{x})^{\dag}T^{\dag}T^{{\dag}*}=\psi(t,\vec{x})^{\dag}T^{\dag}(T^*)^{\dag}=\psi(t,\vec{x})^{\dag}(T^*T)^{\dag}=-\psi(t,\vec{x})^{\dag}$$ or, equivalently, $${\bf T}^{\dag 2}\psi(t,\vec{x}){\bf T}^2=-\psi(t,\vec{x}), \eqno{(30)}$$ consistent with $\psi_{\tau^2}=-\psi$.

\

\

{\bf 6. Compatibility between} $C$ {\bf and} $T$. $CPT$

\

For the time reversal of the Dirac conjugate spinor one has (with $x=(t,\vec{x})$ and $x_\tau=(-t,\vec{x})$): $$\bar{\psi}_\tau(x)=\psi_\tau(x)^{\dag}\gamma_0=(T\psi(x_\tau)^*)^{\dag}\gamma_0=\psi(x_\tau)^{*\dag}T^{\dag}\gamma_0=\psi(x_\tau)^\sim \gamma_0 T^{\dag}=\bar{\psi}(x_\tau)^*T^{\dag};$$
then the time reversal of the charge conjugate spinor is $$(\psi_C)_\tau(x)=C\bar{\psi}_\tau (x)^\sim=C(\psi(x_\tau)^\sim\gamma_0T^{\dag})^\sim=CT^{{\dag}\sim}\gamma_0\psi(x_\tau)=CT^*\gamma_0\psi(x_\tau);$$ on the other hand $$(\psi_C)_\tau(x)=T\psi_C(x_\tau)^*=T(C\bar{\psi}(x_\tau)^\sim)^*=T(C(\psi(x_\tau)^{\dag}\gamma_0)^\sim)^*=T(C\gamma_0\psi(x_\tau)^*)^*=TC^*\gamma_0\psi(x_\tau).$$ Then $C$ and $T$ must be related by $$CT^*=TC^*. \eqno{(31)}$$

\

Consider again the two solutions for $C$: $$i) \ C^*=-C, \ then \ CT^*=-TC \Longleftrightarrow \gamma^2\gamma_0e^{-i\lambda}\gamma^3\gamma^1=e^{-i\lambda}\gamma^3\gamma^1\gamma^2\gamma_0=-e^{i\lambda}\gamma^3\gamma^1\gamma^2\gamma_0$$ which implies  $e^{2i\lambda}=-1$ {\it i.e.} $\lambda=(2k+1){{\pi}\over {2}}$ with $k\in \Z$; then $e^{i\lambda}=(-1)^k i=\{\matrix{i,& k \ even \cr -i, & k \ odd \cr}\}$ and so $$T=\pm i\gamma^3\gamma^1=T^{\dag}=-T^*=T^{-1}=-T^\sim, \ T^2=1. \eqno{(32)}$$  $$ii) \ C^*=C, \ then \ CT^*=TC \Longleftrightarrow \gamma^2\gamma_0e^{-i\lambda}\gamma^3\gamma^1=e^{-i\lambda}\gamma^3\gamma^1\gamma^2\gamma_0=e^{i\lambda}\gamma^3\gamma^1\gamma^2\gamma_0$$ which implies $e^{2i\lambda}=1$ {\it i.e.} $\lambda=k\pi$ with $k\in \Z$; then $e^{i\lambda}=(-1)^k=\{\matrix{1,& k \ even \cr -1, & k \ odd \cr}\}$ and so $$T=\pm \gamma^3\gamma^1=-T^{\dag}=T^*=-T^{-1}=-T^\sim, \ T^2=-1. \eqno{(33)}$$ It is easy to verify that the consistency between $P$ and $T$ does not introduce any additional constraint. In fact, (26) implies $\psi_\tau(t,-\vec{x})=T\psi(-t,-\vec{x})^*$ and the first equality in (11) implies $\psi_\Pi(t,-\vec{x})=P\psi(t,\vec{x})$; then $(\psi_\tau)_\Pi(t,-\vec{x})=P\psi_\tau(t,\vec{x})=PT\psi(-t,\vec{x})^*$ and $(\psi_\Pi)_\tau(t,-\vec{x})=T(\psi_\Pi(-t,-\vec{x}))^*=T(P\psi(-t,\vec{x}))^*=TP^*\psi(-t,\vec{x})^*=-TP\psi(-t,\vec{x})^*$. Since $P\sim \gamma_0$ and $T\sim \gamma^3\gamma^1$, then $PT=TP$ and therefore $$(\psi_\tau)_\Pi=-(\psi_\Pi)_\tau. \eqno{(34)}$$ This equation can be verified at the level of the quantum operators in quantum field theory: from (11), $${\bf T}^{\dag}({\bf P}^{\dag}\psi(t,\vec{x}){\bf P}){\bf T}=({\bf P}{\bf T})^{\dag}\psi(t,\vec{x})({\bf P}{\bf T})={\bf T}^{\dag}P\psi(t,-\vec{x}){\bf T}=P^*{\bf T}^{\dag}\psi(t,-\vec{x}){\bf T}=-P{\bf T}^{\dag} \psi(t,-\vec{x}){\bf T}$$ and, on the other hand, $${\bf P}^{\dag}({\bf T}^{\dag}\psi(t,\vec{x}){\bf T}){\bf P}=({\bf T}{\bf P})^{\dag}\psi(t,\vec{x})({\bf T}{\bf P})=P({\bf T}^{\dag}\psi(t,-\vec{x}){\bf T});$$
then $$({\bf P}{\bf T})^{\dag}\psi(t,\vec{x})({\bf P}{\bf T})=-({\bf T}{\bf P})^{\dag}\psi(t,\vec{x})({\bf T}{\bf P}) \eqno{(35)}$$ which agrees with (34). 

\

In summary, there are only {\bf two sets} of consistent solutions for the matrices $C$, $P$ and $T$ in the case of spin-${{1}\over {2}}$ Dirac fields: $$i) \ C=\pm \gamma^2\gamma_0, \ P=\pm i\gamma_0, \ T=\pm i\gamma^3\gamma^1 \eqno{(36)}$$ with $$C^2=1, \ P^2=-1, \ T^2=1 \eqno{(36a)}$$ and $$ii) \ C=\pm i\gamma^2\gamma_0, \ P=\pm i\gamma_0, \ T=\pm \gamma^3\gamma^1 \eqno{(37)}$$ with $$C^2=P^2=T^2=-1. \eqno{(37a)}$$

\

We notice that the only difference between the two solutions is the $i$ factor interchanged between the matrices $C$ and $T$, what is responsible for the opposite sign in the square of these matrices. Correspondingly, as it will be shown in the following section, there are two $CPT$-groups of sixteen elements each, subgroups of the Dirac algebra. The product matrix $$\theta=CPT \eqno{(38)}$$ is however the {\it same} for the two sets and is given by $$\theta=(\pm\gamma^2\gamma_0)(\pm i\gamma_0)(\pm i\gamma^3\gamma_1)=(\pm i\gamma^2\gamma_0)(\pm i\gamma_0)(\pm \gamma^3\gamma^1)=
\pm \gamma^1 \gamma^2\gamma^3=\pm i \gamma_0\gamma_5=\pm i\pmatrix{1 & 0 \cr 0 & -1 \cr}\pmatrix{0 & -1 \cr -1 & 0}$$ $$=\pm i \pmatrix{0 & -1 \cr 1 & 0 \cr}=\pm \pmatrix{0 & -i1 \cr i1 & 0 \cr} \eqno {(39)}$$ which implies $$\theta^2=-\gamma_0\gamma_5\gamma_0\gamma_5=\gamma_0^2\gamma_5^2=1, \ \theta^{\dag}=\theta=\theta^{-1}=-\theta^\sim =-\theta^*,$$ $$ det(\theta)=1, \ tr(\theta)=0. \eqno {(40)}$$

\

One has $$C, \ T, \ \theta \in {\cal K} \ and \ P\in {\cal M} \eqno{(41)}$$ for case i), and $$ \theta \in {\cal K}, \ P\in {\cal M} \ and \ C, \ T \in {\cal N} \eqno{(42)}$$ for case ii), where ${\cal K}=(SU(4)\cap H(4))_{0,a}$, ${\cal M}= (SU(4)\cap \bar{H}(4))_{0,s}$, and ${\cal N}=(SU(4)\cap \bar{H}(4))_{0,a}$ are respectively the sets of traceless (hemitian, antihermitian, antihermitian) unitary unimodular (antisymmetric, symmetric, antisymmetric) $4\times 4$ complex matrices. Then, in ${\cal K}$ and ${\cal M}$ the matrices are pure imaginary, and in ${\cal N}$ they are real. 

\

\

{\bf 7. Group structures}

\

7.1 {\it Matrix groups}

\

For definiteness, we choose the plus signs in (36) and (37), respectively obtaining $$P=i\gamma_0, \ C=\gamma^2\gamma_0, \ and \ \ T=i\gamma^3\gamma^1, \eqno{(36b)}$$ and $$P=i\gamma_0, \ C=i\gamma^2 \gamma_0, \ and \ \ T=\gamma^3 \gamma^1. \eqno{(37b)}$$ By taking products, these matrices generate two non abelian groups of sixteen elements each, respectively $G_\theta^{(1)}$ and $G_\theta^{(2)}$, which are subgroups of the Dirac algebra with the inherited algebra multiplication: $$\{\pm 1,\pm C,\pm P,\pm T,\pm CP, \pm CT, \pm PT, \pm \theta \}.$$  Clearly, the choice of signs in (36b) and (37b) (among 2$^3$=8 possibilities in each case) does not alter the structure of these groups. 

\

Their basic multiplication tables are the following: 

\

$G_\theta^{(1)}$:

\

$$\matrix{&C&P&T&CP&CT&PT&\theta&\cr C&1&CP&CT&P&T&\theta&PT&\cr P&-CP&-1&PT&C&-\theta&-T&CT&\cr T&CT&PT&1&\theta&C&P&CP&\cr CP&-P&-C&\theta&1&-PT&-CT&T&\cr
CT&T&\theta&C&PT&1&CP&P&\cr PT&-\theta&-T&P&CT&-CP&-1&C&\cr \theta&-PT&-CT&CP&T&-P&-C&1&\cr} \eqno{(43)}$$ 

\

$G_\theta^{(2)}$:

\

$$\matrix{&C&P&T&CP&CT&PT&\theta& \cr C&-1&CP&CT&-P&-T&\theta&-PT&\cr P&-CP&-1&PT&C&-\theta&-T&CT&\cr T&CT&PT&-1&\theta&-C&-P&-CP&\cr CP&P&-C&\theta&-1&PT&-CT&-T&\cr CT&-T&\theta&-C&-PT&1&-CP&P&\cr PT&-\theta&-T&-P&CT&CP&1&-C&\cr \theta&PT&-CT&-CP&-T&-P&C&1&\cr}\eqno{(44)}$$ The tables are completed by adding to the first row and to the first column of each table, the negatives $-C, \ -P, \ ..., \ -\theta$ and -1, and making the corresponding products; then one obtains identical diagonal blocks and their negatives for the non diagonal blocks. 

$G_\theta^{(1)}$ has eleven elements of order 2: $\{-1,\pm C,\pm T, \pm CP, \pm CT, \pm \theta\}$ (the identity is of order 1), and four elements of order 4: $\{\pm P, \pm PT\}$; and $G_\theta^{(2)}$ has seven elements of order 2: $\{-1,\pm CT, \pm PT, \pm \theta\}$, and eight elements of order 4: $\{\pm C, \pm P, \pm T, \pm CP\}$.  

By Cayley theorem, $G_\theta^{(1)}$ and $G_\theta^{(2)}$ are isomorphic to regular subgroups of $S_{16}$, the {\it symmetric group} of 16 elements (Hamermesh, 1989), of order 16!=2.0922789888$\times 10^{13}$. (A regular subgroup of $S_n$ is a subgroup of order $n$ consisting of permutations which do not leave any symbol in $\{1,2,...,n\}$ unchanged, except for the identity in which no symbol is changed.) Labelling the elements $$1, \ C, \ P, \ T, \ CP, \ CT, \ PT, \ \theta, \ -C, \ -P, \ -T, \ -CP, \ -CT, \ -PT, \ -\theta \ and \ -1$$ respectively by 1, 2, 3, 4, 5, 6, 7, 8, 9, 10, 11, 12, 13, 14, 15 and 16, the permutations corresponding to $1, \ C, \ P, \ T, \ ..., \ -\theta, \ -1$, expressed in terms of their cycles, are the following:

\

For $G_\theta^{(1)}$:

\ 

$1 \ \leftrightarrow$ (1) (2)...(16),

$C \ \leftrightarrow$ (1 2) (3 5) (4 6) (7 8) (9 16) (10 12) (11 13) (14 15),

$P \ \leftrightarrow$ (1 3 16 10) (2 12 9 5) (4 7 11 14) (6 15 13 8),

$T \ \leftrightarrow$ (1 4) (2 6) (3 7) (5 8) (9 13) (10 14) (11 16) (12 15),

$CP \ \leftrightarrow$ (1 5) (2 10) (3 9) (4 8) (6 14) (7 13) (11 15) (12 16),

$CT \ \leftrightarrow$ (1 6) (2 4) (3 8) (5 7) (9 11) (10 15) (12 14) (13 16),

$PT \ \leftrightarrow$ (1 7 16 14) (2 15 9 8) (3 11 10 4) (5 6 12 13),

$\theta \ \leftrightarrow$ (1 8) (2 14) (3 13) (4 5) (6 10) (7 9) (11 12) (15 16),

$-C \ \leftrightarrow$ (1 9) (2 16) (3 12) (4 13) (5 10) (6 11) (7 15) (8 14),

$-P \ \leftrightarrow$ (1 10 16 3) (2 5 9 12) (6 8 13 15) (4 14 11 7),

$-T \ \leftrightarrow$ (1 11) (2 13) (3 14) (4 16) (5 15) (6 9) (7 10) (8 12),

$-CP \ \leftrightarrow$ (1 12) (2 3) (4 15) (5 16) (6 7) (8 11) (9 10) (13 14),

$-CT \ \leftrightarrow$ (1 13) (2 11) (3 15) (4 9) (5 14) (6 16) (7 12) (8 10),

$-PT \ \leftrightarrow$ (1 14 16 7) (2 8 9 15) (3 4 10 11) (12 6 5 13),

$-\theta \ \leftrightarrow$ (1 15) (2 7) (3 6) (4 12) (5 11) (8 16) (9 14) (10 13),

$-1 \ \leftrightarrow$ (1 16) (2 9) (3 10) (4 11) (5 12) (6 13) (7 14) (8 15). $$\eqno{(45)}$$ {\it I.e.} the identity decomposes into sixteen cycles of length 1, eleven elements decompose into eight cycles of length 2, and four elements decompose into four cycles of length 4.
 
\

For $G_\theta^{(2)}$:

\

$1 \ \leftrightarrow$ (1) (2)...(16),

$C \ \leftrightarrow$ (1 2 16 9) (3 5 10 12) (6 11 13 4) (7 8 14 15),

$P \ \leftrightarrow$ (1 3 16 10) (2 12 9 5) (4 7 11 14) (6 15 13 8),

$T \ \leftrightarrow$ (1 4 16 11) (2 6 9 13) (5 8 12 15) (3 7 10 14),

$CP \ \leftrightarrow$ (1 5 16 12) (2 3 9 10) (4 8 11 15) (6 7 13 14),

$CT \ \leftrightarrow$ (1 6) (2 11) (3 8) (4 9) (5 14) (7 12) (10 15) (13 16),

$PT \ \leftrightarrow$ (1 7) (16 14) (2 15) (3 11) (4 10) (8 9) (12 13) (5 6),

$\theta \ \leftrightarrow$ (1 8) (2 7) (3 13) (4 12) (5 11) (6 10) (9 14) (15 16),

$-C \ \leftrightarrow$ (1 9 16 2) (3 12 10 5) (4 13 11 6) (7 15 14 8),

$-P \ \leftrightarrow$ (1 10 16 3) (2 5 9 12) (4 14 11 7) (6 8 13 15),

$-T \ \leftrightarrow$ (1 11 16 4) (2 13 9 6) (5 15 12 8) (7 3 14 10),

$-CP \ \leftrightarrow$ (1 12 16 5) (2 10 9 3) (4 15 11 8) (6 14 13 7),

$-CT \ \leftrightarrow$ (1 13) (2 4) (3 15) (5 7) (6 16) (8 10) (9 11) (12 14),

$-PT \ \leftrightarrow$ (1 14) (2 8) (3 4) (5 13) (6 12) (7 16) (9 15) (10 11),

$-\theta \ \leftrightarrow$ (1 15) (2 14) (3 6) (4 5) (7 9) (8 16) (10 13) (11 12),

$-1 \ \leftrightarrow$ (1 16) (2 9) (3 10) (4 11) (5 12) (6 13) (7 14) (8 15). $$\eqno{(46)}$$ {\it I. e.}, the identity decomposes into sixteen cycles of length 1, seven elements decompose into eight cycles of length 2, and eight elements decompose into four cycles of length 4.

\

Moreover, $G_\theta^{(1)}$ and $G_\theta^{(2)}$ turn out to be finite subgroups of the $Pin$ group (Lawson and Michelsohn, 1989) of the Dirac algebra, $Pin_{D^{16}}$. In fact, for $\sigma$=1 and 2 one has the sequence of groups and group inclusions (homomorphisms): 
$$G_\theta^{(\sigma)}\to Pin_{D^{16}}\to P_{D^{16}} \to Cl_{D^{16}} \to D^{16*} \to D^{16}, \eqno{(47)}$$ where:

\

$D^{16*}$: {\it group of units} of $D^{16}$, namely the set of elements of $D^{16}$ with multiplicative inverse; $D^{16*}$ is a Lie group with Lie algebra $Lie(D^{16*})=d^{16}$ (=$D^{16}$ as a set) equiped with the Lie bracket $[v_1,v_2]=v_1v_2-v_2v_1$. The {\it adjoint representation} and the {\it twisted adjoint representation} of $D^{16*}$, which are representations of $D^{16*}$ over $d^{16}$, are respectively the group homomorphisms $Ad:D^{16*}\to GL(d^{16})$ and $\tilde{Ad}:D^{16*}\to GL(d^{16})$ given by $Ad(v)(w)=vwv^{-1}$ and $\tilde{Ad}(v)(w)=\alpha(v)wv^{-1}$, where $\alpha:D^{16}\to D^{16}$, the {\it canonical involution} of $D^{16}$, is the map of algebras induced by $x^\mu \to -x^\mu=x^\mu_{\Pi \tau}$ in $M^4$ or, equivalently, by $\iota (M^4) \to \iota(M^4)$ in $D^{16}$ given by $x^\mu \gamma_\mu \to -x^\mu \gamma_\mu$, where $\iota:M^4 \to D^{16}$ is the {\it canonical inclusion} $\iota (x^\mu)=x^\mu \gamma_\mu$. For example, if $x^\mu \gamma_\mu \in \iota(M^4)$ then for all $y^\nu \gamma_\nu \in \iota(M^4)$, $-Ad_{x^\mu \gamma_\mu}(y^\nu \gamma_\nu)=y^\nu \gamma_\nu-{{2\eta_{\rho\sigma}x^\rho y^\sigma}\over {x^2}}x^\nu \gamma_\nu$ is the reflection of $\iota(y^\mu)$ across the hyperplane perpendicular to $\iota(x^\mu)$, and $\tilde{Ad}_{x^\mu \gamma_\mu}(y^\nu \gamma_\nu)=-Ad_{x^\mu \gamma_\mu}(y^\nu \gamma_\nu)$.

\

$Cl_{D^{16}}$: {\it Clifford group} of $D^{16}$: $\{v\in D^{16*}\vert \tilde{Ad}(v)(\iota(M^4)_c)\subset \iota(M^4)_c \}$ where $\iota(M^4)_c=\{z^\mu \gamma_\mu, \ z^\mu \in \C \}$ is the complexification of $\iota(M^4)$; clearly, $\iota(M^4)\cong \C ^4$.

\

$P_{D^{16}}=\{v_1 \cdots v_p \vert v_k \in (\iota(M^4)_c)^*=\iota(M^4)_c\cap D^{16*}, \ k=1,\cdots,p, \ p \in \Z^+\}$; $(v_1 \cdots v_p)^{-1}=v_p^{-1}\cdots v_1^{-1}$, $v_k^{-1}=(z^\mu \gamma_\mu)^{-1}={{z^\mu}\over{z^\nu z_\nu}}\gamma_\mu$, $z^\nu z_\nu\neq 0$; and $(\iota(M^4))^*=\hat{q}^{-1}(\R^*)$ where $\hat{q}:\iota(M^4)\to \R$ is given by $\hat{q}(\iota \vert (x^\mu))=\hat{q}\circ \iota \vert (x^\mu)= q(x^\mu)=\eta_{\mu \nu}x^\mu x^\nu$ (or $\tilde{\eta}_{\mu \nu}x^\mu x^\nu$) with $\iota \vert: M^4\to \iota (M^4), \ \iota \vert (x^\mu)=\iota (x^\mu)$. Clearly, $P_{D{16}}\subset Cl_{D^{16}}$ since $\tilde{Ad}_{v_1\cdots v_p}(x^\mu \gamma_\mu)=\alpha(v_1\cdots v_p)x^\mu \gamma_\mu v_p^{-1}\cdots v_1^{-1}=\alpha(v_1)\cdots (\alpha(v_p)x^\mu \gamma_\mu v_p^{-1})\cdots v_1^{-1}=y^\mu \gamma_\mu$. 

\

$Pin_{D^{16}}=\{v_1\cdots v_p \vert v_k \in (\hat{q}^{-1}(\{1,-1\}))_c \}$. 

\

It is clear that $$G_\theta ^{(\sigma)}\subset Pin_{D^{16}}, \ \ \sigma=1,2; \eqno{(48)}$$ however, since $G_{\theta}^{(\sigma)}$ contains even and odd elements of $D^{16}$, then $G_{\theta}^{(\sigma)} \not\subset Spin_{D^{16}}=Pin_{D^{16}} \cap D^{16}_+$. ($D^{16}_+$ ($D^{16}_-$) is the even (odd) part of the Dirac algebra in the direct sum decomposition $D^{16}=D^{16}_+ \oplus D^{16}_-$.) In particular, then, the groups $G_{\theta}^{(\sigma)}$ are not contained in the connected component of $Spin_{D^{16}}$, $Spin^0_{D^{16}} \cong SL(2, \C)\oplus SL(2,\C)$, the universal covering group of the connected component ${\cal L}_{c+}$ of the complex Lorentz group ${\cal L}_c$.  

\

We now present a more detailed investigation of the group structures of $G_\theta^{(1)}$ and $G_\theta^{(2)}$, and the geometrical elements which are involved. As a consequence of Cayley theorem, for each positive integer $n$ the number of groups with $n$ elements is finite. In particular, there exist fourteen groups of sixteen elements, see e.g. (Asche, 1989): five of them are abelian and the remaining nine are non abelian; from these, only three have 3 generators: $DH_8 \times \Z_2$, $DC_8\times \Z_2$, and $16E$; here, $DH_8$ is the dihedral group of eight elements, $DC_8$ is the dicyclic group of eight elements, and $16E$ is an extension of $DH_8$ by an element of order 4 (see below). As a subgroup of $S_8$, the generators of $DC_8$ can be chosen as $x=(1234)(5678)$ and $y=(1537)(2846)$, then $DC_8=<\{x,y\}>=\{1,x,x^2,x^3,y,xy,x^2y,x^3y\}$ with $x^2$ of order 2, and $x$, $x^3$, $y$, $xy$, $x^2y$ and $x^3y$ of order 4. Then the direct product $DC_8\times \Z_2\subset S_{10}$ has three elements of order 2: $(x^2,1)$, $(x^2,z)$ and $(1,z)$ with $z=(9 \ 10)$, and twelve elements of order 4: $(x,1)$, $(x,z)$, $(x^3,1)$, $(x^3,z)$, $(y,1)$, $(y,z)$, $(xy,1)$, $(xy,z)$, $(x^2y,1)$, $(x^2y,z)$, $(x^3y,1)$ and $(x^3y,z)$. So, neither $G_\theta^{(1)}$ nor $G_\theta^{(2)}$ is isomorphic to $DC_8\times\Z_2$. The remaining candidates are $DH_8\times \Z_2$ and $16E$, and we find the isomorphisms $$G_\theta^{(1)}\cong DH_8\times \Z_2 \ \ and \ \ G_\theta^{(2)}\cong 16E. \eqno{(49)}$$

\

$DH_8$, the {\it symmetry group of the square}, consists of four rotations: $1=0^\circ$, $r=90^\circ$, $r^2=180^\circ$ and $r^3=270^\circ$, and four reflections: two in the diagonals and two in the axis joining the midpoints of opposite edges. Identifying $r=(1234)$ and the reflection $b=(24)$ we obtain $$DH_8=<\{r,b\}>=\{1,r,r^2,r^3,b,rb,r^2b,r^3b\}$$ $$=\{1,(1234),(13)(24),(1432),(24),(12)(34),(13),(14)(23)\}\subset S_4. \eqno{(50)}$$ Then $$DH_8\times \Z_2 \subset S_6, \eqno{(51)}$$ the {\it product (trivial) extension} of $DH_8$ by $\Z_2= \{1,(56)\}$, $DH_8\times \Z_2=<\{(1234),(24),(56)\}>$, has elements $$DH_8\times\Z_2=\{1,(1234),(13)(24),(1432),(24),(12)(34),(13),(14)(23),(56),(1234)(56),(13)(24)(56),$$ $$(1432)(56),(24)(56),(12)(34)(56),(13)(56),(14)(23)(56)\}, \eqno{(52)}$$ of which eleven are of order 2: (24), (13), (56), (13)(24), (13)(24)(56), (24)(56), (12)(34), (12)(34)(56), (13)(56), (14)(23) and (14)(23)(56), and four are of order 4: (1234), (1234)(56), (1432) and (1432)(56). The {\it isomorphism} between $G_\theta^{(1)}$ and $DH_8\times \Z_2$, $$\Psi^{(1)}:G_\theta^{(1)}\to DH_8\times \Z_2,$$ as can be verified after a straightforward calculation, is given by: 

\

$1 \longmapsto 1$

$C \longmapsto (24)$

$P \longmapsto (1234)$

$T \longmapsto (56)$

$CP \longmapsto (14)(23)$

$CT \longmapsto (24)(56)$

$PT \longmapsto (1234)(56)$

$\theta \longmapsto (14)(23)(56)$

$-1 \longmapsto (13)(24)$

$-C \longmapsto (13)$

$-P \longmapsto (1432)$

$-T \longmapsto (13)(24)(56)$

$-CP \longmapsto (12)(34)$

$-CT \longmapsto (13)(56)$

$-PT \longmapsto (1432)(56)$

$-\theta \longmapsto (12)(34)(56).$  $$\eqno{(53)}$$

Also, one has the short exact sequence of groups and group homomorphisms (Mac Lane and Birkoff, 1979) $$0 \to DH_8 \buildrel {\iota_1} \over \longrightarrow DH_8\times \Z_2 \buildrel {\varphi_1} \over \longrightarrow \Z_2\to 0, \eqno{(54)}$$ with $\iota_1(g)=(g,1)$ and $\varphi_1(g,h)=h$; the sequence splits through the group homomorphism $\gamma_1:\Z_2\to DH_8\times \Z_2$, $\gamma_1(h)=(1,h)$ {\it i.e.} $\varphi_1 \circ \gamma_1=Id_{\Z_2}$. 

\

The group $16E\subset S_8$ is generated by $a=(1234)(5678)$, $d=(1638)(2547)$ and $n=(17)(28)(35)(46)$. Then, it can be easily verified that $a^2=d^2=-1$, $n$, $a^2n$, $dn$, $nd$, $and$ and $adn$ have order 2, and the eight elements $a$, $a^3$, $d$, $d^3$, $an$, $ad$, $da$ and $a^3n$ have order 4. Then $$\psi^{(2)}:G_\theta^{(2)}\to 16E$$ given by:

\

$1 \longmapsto 1$

$C \longmapsto a$

$P \longmapsto d$

$T \longmapsto an=na=(1836)(2547)$

$CP \longmapsto ad=(1735)(2648)$

$CT \longmapsto a^2n=-n=(15)(26)(37)(48)$

$PT \longmapsto and=-adn=(24)(57)$

$\theta \longmapsto dn=(12)(34)(58)(67)$

$-1 \longmapsto -1=(13)(24)(57)(68)$

$-C \longmapsto a^3=-an=(1432)(5876)$

$-P \longmapsto d^3=-d=(1836)(2745)$

$-T \longmapsto a^3n=-an=(1638)(2745)$

$-CP \longmapsto da=-ad=(1537)(2846)$

$-CT \longmapsto n$

$-PT \longmapsto adn=(13)(68)$

$-\theta \longmapsto nd=-dn=(14)(23)(56)(78)$ $$\eqno{(55)}$$ is an {\it isomorphism} between $G_\theta^{(2)}$ and $16E$. 

\

On the other hand, one can verify that the subgroup of $16E$ generated by $d$ and $n$, namely $$\{1,-1,d,-d,n,-n,dn,-dn\},$$ is isomorphic to $DH_8$, which, having index 2 in $16E$, is an invariant subgroup {\it i.e.} $gDH_8g^{-1}=DH_8$ for all $g\in 16E$. (This can be easily verified by an explicit  calculation.) The isomorphism $<\{d,n\}>\to DH_8$ is given by $d\to (1234)$ and $n \to (24)$. Then one has the short exact sequence $$0 \to DH_8 \buildrel {\iota_2} \over \longrightarrow 16E \buildrel {\varphi_2} \over \longrightarrow \Z_2 \to 0 \eqno{(56)}$$ (since ${{16E}\over {DH_8}}\cong \Z_2$) with $\iota_2$ the inclusion, $Ker(\varphi_2)=DH_8$, and $\varphi_2(a)=\varphi_2(an)=\varphi_2(ad)=\varphi_2(and)=\varphi_2(a^3)=\varphi_2(a^3n)=\varphi_2(da)=\varphi_2(adn)=-1$. In other words, $16E$ -and therefore $G_\theta^{(2)}$- is also an {\it extension} though {\it not the trivial one} of $DH_8$ by $\Z_2$. The extension {\it splits}, that is, there is a group homomorphism $\gamma_2: \Z_2 \to 16E$ given by $\gamma_2(1)=1$ and $\gamma_2(-1)=adn$ (or $\gamma_2(-1)=and$) with $\varphi_2 \circ \gamma_2=Id_{\Z_2}$. Let us choose $\gamma_2(-1)=adn$; then there is the {\it isomorphism} $$\psi_2:DH_8\times _{\Phi_2} \gamma_2(\Z_2)\to 16E, \ \ \psi_2(g,\gamma_2(h))=g\gamma_2(h), \eqno{(57)}$$ where the composition in the semidirect product $DH_8\times _{\Phi_2}\gamma_2(\Z_2)$ is $$(g^\prime,\gamma_2(h^\prime))(g,\gamma_2(h))=(g^\prime \Phi_2(h^\prime)(g),\gamma_2(h^\prime)\gamma_2(h)) \eqno{(58)}$$ with $\Phi_2:\Z_2\to Aut(DH_8)$ given by $\Phi_2(h^\prime)(g)=\gamma_2(h^\prime)g \gamma_2(h^\prime)^{-1}$. Explicitly, the isomorphism $\psi_2$ is given by:

\

$(1,1)\longmapsto 1$

$(1,adn)\longmapsto adn$

$(n,1)\longmapsto n$

$(n,adn)\longmapsto da$

$(-1,1)\longmapsto -1$

$(-1,adn)\longmapsto -adn$

$(-n,1)\longmapsto -n$

$(-n,adn)\longmapsto -da$

$(d,1)\longmapsto d$

$(d,adn)\longmapsto an$

$(dn,1)\longmapsto dn$

$(dn,adn)\longmapsto -a$

$(-d,1)\longmapsto -d$

$(-d,adn)\longmapsto -an$

$(-dn,1)\longmapsto -dn$

$(-dn,adn)\longmapsto a$,  $$\eqno{(59)}$$ and therefore for the composition $$\Psi^{(2)}\equiv \psi_2^{-1}\circ\psi^{(2)}:G_\theta^{(2)}\to DH_8\times_{\Phi_2}\gamma_2(\Z_2)$$ one has

\

$1 \longmapsto (1,1)$

$C \longmapsto (-dn,adn)$

$P \longmapsto (d,1)$

$T \longmapsto (d,adn)$

$CP \longmapsto (-n,adn)$

$CT \longmapsto (-n,1)$

$PT \longmapsto (-1,adn)$

$\theta \longmapsto (dn,1)$

$-1 \longmapsto (-1,1)$

$-C \longmapsto (dn,adn)$

$-P \longmapsto (-d,1)$

$-T \longmapsto (-d,adn)$

$-CP \longmapsto (n,adn)$

$-CT \longmapsto (n,1)$

$-PT \longmapsto (1,adn)$

$-\theta \longmapsto (-dn,1)$. $$\eqno{(60)}$$

\

Notice that since $DH_8$ is not abelian, both $G_\theta^{(1)}$ and $G_\theta^{(2)}$ are {\it non central} and {\it non abelian} extensions of $DH_8$ by $\Z_2$. 

\

$DH_8$ and $\Z_2$ have a natural geometric content since, as we said before, $DH_8$ consists of the eight symmetries of the square, and $\Z_2 \cong S^0$, the 0-sphere. We can however go into a more elementary description of the $CPT$ groups by noticing that $\Z_4=\{1,d,-1,-d\}$ is an invariant subgroup of $DH_8$, with the quotient ${{DH_8}\over {\Z4}}\cong \Z_2$. Then one has the short exact sequence $$0\to \Z_4 \buildrel {\iota} \over \longrightarrow DH_8 \buildrel {\varphi} \over \longrightarrow \Z_2 \to 0 \eqno{(61)}$$ where $\iota$ is the inclusion, $Ker(\varphi)=\Z_4$, and $\varphi(n)=\varphi(dn)=\varphi(-n)=\varphi(-dn)=-1$. Then $DH_8$ is an {\it abelian non trivial extension} of $\Z_4$ by $\Z_2$; the extension is {\it non central} since the center of $DH_8$ is $\{1,-1\}$ and $\Z_4 \not \subset \{1,-1\}$. The extension splits through $\gamma:\Z_2 \to DH_8$ given by $\gamma(1)=1$ and $\gamma(-1)=n$ or $\gamma(-1)=dn$, with $\varphi \circ \gamma=Id_{\Z_2}$. Let us choose $\gamma(-1)=n$; then one has the group isomorphism $$\psi:\Z_4\times_{\Phi}\gamma(\Z_2)\to DH_8, \ \ \psi(g,\gamma(h))=g\gamma(h), \eqno{(62)}$$ where the composition in the semidirect product is $(g^\prime,\gamma(h^\prime))(g,\gamma(h))=(g^\prime\Phi (h^\prime)(g),\gamma(h^\prime)\gamma(h))$ with $\Phi(h^\prime)\in Aut(\Z_4)$ given by $\Phi(h^\prime)(g)=\gamma(h^\prime)g\gamma(h^\prime)^{-1}$. For the isomorphism $\psi$ one has:

\

$(1,1)\longmapsto 1$

$(1,n)\longmapsto n$

$(-1,1)\longmapsto -1$

$(-1,n)\longmapsto -n$

$(d,1)\longmapsto d$

$(d,n)\longmapsto dn$

$(-d,1)\longmapsto -d$

$(-d,n)\longmapsto -dn$.  $$\eqno{(63)}$$ 

\

In summary, the group structures of $G_{\theta}^{(1)}$ and $G_\theta^{(2)}$ suggest a {\it geometrical nature} of the three discrete operations, $C$, $P$ and $T$, and of their product $\theta$, besides the one associated with the fact that $\Pi$ and $\tau$ are elements of the Lorentz group ${\cal L}$. In particular, this is relevant for the charge conjugation operation $c$, which, as we mentioned in the introduction, does not sit in the Poincar\'e group. The crucial point is that, on the one hand,  $G_{\theta}^{(1)}$ and $G_\theta^{(2)}$ are subgroups of a Clifford algebra ($D^{16}$), which besides being a geometrical algebra, because it is determined by a metric in a vector space (Porteous, 1981), it is the universal object of a certain functor (Aguilar and Socolovsky, 1997); on the other hand, the short exact sequences (54), (56) and (61) show that, in the last instance, $G_\theta^{(1)}$ and $G_\theta^{(2)}$ are determined by the groups of the 4-th ($\Z_4$) and the square ($\Z_2$) roots of unity. 

\

In the next subsection, however, we shall show that the requirement of consistency between the one particle Dirac theory and the quantum field theory, selects the second solution, that is the group $G_\theta^{(2)}$.

\

Another approaches to a geometrical interpretation of charge conjugation are considered by Azc\'arraga and Boya (Azc\'arraga and Boya, 1975); Sternberg (Sternberg, 1987); S\'anchez Valenzuela (S\'anchez Valenzuela, 1991); Shirokov (Shirokov, 1958); and Varlamov (Varlamov, 2003). A general review of the $C$, $P$, $T$ and $\theta$ transformations can be found in (Azc\'arraga, 1975).

\

7.2 {\it Operator group}

\

Even if the groups $G_\theta^{(1)}$ and $G_\theta^{(2)}$ have interesting geometric properties, there is no {\it a priori} physical or mathematical reason to prefer one group to the other in what might be called the $CPT$ group of the Dirac field. 

Instead, the quantum operators $\bf C$, $\bf P$ and $\bf T$, which act on the Hilbert space of the field theory and transform the field operator $\psi(t,\vec{x})$ according to the equations (22), (11) and (29) respectively, are the generators of a unique group $G_\Theta$, which we call {\it the} $\bf CPT$ {\it group of the Dirac field}. $G_\Theta$ has sixteen elements, it is non abelian, and it is isomorphic to the direct product $DC_8\times \Z_2$, where $DC_8$ is the dicyclic group of eight elements, already discussed in the paragraph preceding equation (49). As will be shown below, $G_{\Theta}$ selects $G_{\theta}^{(2)}$ as {\it the matrix} $CPT$ {\it group} of the Dirac field. It is interesting to notice, however, that $G_\theta^{(1)}$, $G_\theta^{(2)}$, and $G_\Theta$ exhaust the non abelian groups of sixteen elements having three generators (Asche, 1989). 

\

Let $\psi=\psi(t,\vec{x})$ be the Dirac field operator, and $\bf A$ and $\bf B$ any of the operators $\bf C$, $\bf P$ and $\bf T$. One defines $$\bf A \cdot \psi=\bf A ^{\dag} \psi \bf A \eqno{(64)}$$ and $$(\bf A *\bf B)\cdot \psi=(\bf A \bf B)^{\dag}\psi (\bf A \bf B). \eqno{(65)}$$ In the r.h.s. of (65), $\bf A \bf B$ is the usual (associative) composition of operators. It is then easy to prove that $\bf A * \bf B$ is also an associative product: in fact, $(\bf A *\bf B)\cdot \psi=\bf B ^{\dag}(\bf A^{\dag}\psi \bf A)\bf B=\bf B\cdot(\bf A\cdot \psi)$ and so $((\bf A*\bf B)*\bf C)\cdot \psi=\bf C\cdot((\bf A*\bf B)\cdot \psi)=\bf C\cdot(\bf B\cdot(\bf A\cdot \psi))$ and $(\bf A*(\bf B*\bf C))\cdot \psi=(\bf B*\bf C)\cdot (\bf A\cdot \psi)=\bf C\cdot (\bf B\cdot(\bf A\cdot \psi))$; since this holds for all values of $\psi$, then $$(\bf A*\bf B)*\bf C=\bf A*(\bf B*\bf C). \eqno{(66)}$$ In equations (12a), (23), (30) and (35) we proved, respectively, that $$\bf P*\bf P=-1, \ \bf C*\bf C=1, \ \bf T*\bf T=-1 \ and \ \bf T*\bf P=-\bf P*\bf T. \eqno{(67)}$$ Through a similar calculation one obtains (see Appendix 2) $$\bf C*\bf P=\bf P*\bf C \ \ and \ \ \bf C*\bf T= \bf T*\bf C. \eqno{(68)}$$ 
The second equality together with (31) imply (Appendix 2) $$T^*=T, \eqno{(69)}$$ which {\bf selects the matrix group} $G_\theta^{(2)}$, that is, the solution (37), (37a) or, for signs definiteness, (37b). 

\

For the quantum operators, one has the multiplication table: $$\matrix{&\bf C&\bf P&\bf T\cr \bf C&1&\bf C*\bf P&\bf C*\bf T\cr \bf P&\bf C*\bf P&-1&\bf P*\bf T\cr \bf T&\bf C*\bf T&-\bf P*\bf T&-1\cr} \ \ , \eqno{(70)}$$ from which, using associativity, one obtains the basic multiplication table of the group $G_\Theta$, where $\Theta=\bf C*\bf P*\bf T$: $$\matrix{&\bf C&\bf P&\bf T&\bf C*\bf P&\bf C*\bf T&\bf P*\bf T&\Theta \cr \bf C&1&\bf C*\bf P&\bf C*\bf T&\bf P&\bf T&\Theta &\bf P*\bf T\cr \bf P&\bf C*\bf P&-1&\bf P*\bf T&-\bf C&\Theta &-\bf T&-\bf C*\bf T \cr \bf T&\bf C*\bf T&-\bf P*\bf T&-1&-\Theta &-\bf C&\bf P&\bf C*\bf P \cr \bf C*\bf P&\bf P&-\bf C&\Theta &-1&\bf P*\bf T&-\bf C*\bf T&-\bf T \cr \bf C*\bf T&\bf T&-\Theta &-\bf C&-\bf P*\bf T&-1&\bf C*\bf P&\bf P \cr \bf P*\bf T&\Theta &\bf T&-\bf P&\bf C*\bf T&-\bf C*\bf P&-1&-\bf C \cr \Theta &\bf P*\bf T&\bf C*\bf T&-\bf C*\bf P&\bf T&-\bf P&-\bf C&-1 \cr} \eqno{(71)}$$ The table is completed by adding to the first row and to the first column, the negatives $-\bf C, \ -\bf P, \ \dots, \ -\Theta$, and -1, and making the corresponding products; then one obtains identical diagonal blocks and their negatives for the non diagonal blocks. For the {$\bf P$} and {$\bf T$} transformations, this group structure coincides with that of the group $G_5$ of Shirokov (Shirokov, 1960) for the case of half-integer spins; and with Feynman (Feynman, 1987) and Sakurai (Sakurai, 1985) for the case of the square of the {$\bf T$} operator. 

So, $G_\Theta$ is a non abelian group of sixteen elements, three generators, twelve elements of order 4: $\{\pm \bf P, \ \pm \bf T, \ \pm \bf C*\bf P, \ \pm \bf C*\bf T, \ \pm \bf P*\bf T, \ \pm \Theta \}$, three elements of order 2: $\{\pm \bf C, \ -1\}$, and one element of order 1: $\{1\}$. Then, $$G_\Theta \cong DC_8\times \Z_2 \ . \eqno{(72)}$$ As is well known (Armstrong, 1988), $DC_8$ is isomorphic to the {\it quaternion} group $Q$, generated by the imaginary units $\iota$ and $\gamma$, with the isomorphism given by $x\mapsto \iota$ and $y\mapsto \gamma$. The multiplication table for the three imaginary units, $\iota$, $\gamma$ and $\kappa$ is the following: $$\matrix{&\iota &\gamma &\kappa \cr \iota &-1&\kappa &-\gamma \cr \gamma &-\kappa &-1&\iota \cr \kappa &\gamma &-\iota &-1\cr}$$ Then, as can be verified after a long but straightforward calculation, one has the following sequence of isomorphisms: $$G_\Theta \buildrel {\Psi} \over \longrightarrow DC_8\times \Z_2\longrightarrow Q\times S^0\longrightarrow H\longrightarrow K \ :$$

\noindent $1\mapsto (1,1)\mapsto (1,1)\mapsto 1\mapsto 1$

\noindent ${\bf C} \mapsto (1,z)\mapsto (1,-1)\mapsto (9 \ 10)\mapsto (1 \ 2)(3 \ 5)(4 \ 6)(7 \ 8)(9 \ 16)(10 \ 12)(11 \ 13)(14 \ 15)$

\noindent ${\bf P} \mapsto (x,1)\mapsto (\iota,1)\mapsto (1 2 3 4)(5 6 7 8)\mapsto (1 \ 3 \ 16 \ 10)(2 \ 5 \ 9 \ 12)(4 \ 7 \ 11 \ 14)(6 \ 8 \ 13 \ 15)$

\noindent ${\bf T} \mapsto (y,1)\mapsto (\gamma,1)\mapsto (1 5 3 7)(2 8 4 6)\mapsto (1 \ 4 \ 16 \ 11)(2 \ 6 \ 9 \ 13)(3 \ 14 \ 10 \ 7)(5 \ 15 \ 12 \ 8)$

\noindent ${\bf C*\bf P} \mapsto (x,z)\mapsto (\iota,-1)\mapsto (1 2 3 4)(5 6 7 8)(9 \ 10)\mapsto (1 \ 5 \ 16 \ 12)(2 \ 3 \ 9 \ 10)(4 \ 8 \ 11 \ 15)(6 \ 7 \ 13 \ 14)$

\noindent ${\bf C*\bf T} \mapsto (y,z)\mapsto (\gamma,-1)\mapsto (1537)(2846)(9 \ 10)\mapsto (1 \ 6 \ 16 \ 13)(2 \ 4 \ 9 \ 11)(3 \ 15 \ 10 \ 8)(5 \ 14 \ 12 \ 7)$

\noindent ${\bf P*\bf T} \mapsto (xy,1)\mapsto (\kappa,1)\mapsto (1638)(2547)\mapsto (1 \ 7 \ 16 \ 14)(2 \ 8 \ 9 \ 15)(3 \ 4 \ 10 \ 11)(5 \ 6 \ 12 \ 13)$

\noindent $\Theta \mapsto (xy,z)\mapsto(\kappa,-1)\mapsto (1638)(2547)(9 \ 10)\mapsto (1 \ 8 \ 16 \ 15)(2 \ 7 \ 9 \ 14)(3 \ 6 \ 10 \ 13)(4 \ 12 \ 11 \ 5)$

\noindent $-{\bf C} \mapsto (x^2,z)\mapsto (-1,-1)\mapsto (13)(24)(57)(68)(9 \ 10)\mapsto (1 \ 9)(2 \ 16)(3 \ 12)(4 \ 13)(5 \ 10)(6 \ 11)(7 \ 15)(8 \ 14)$

\noindent $-{\bf P} \mapsto (x^3,1)\mapsto (-\iota,1)\mapsto (1432)(5876)\mapsto (1 \ 10 \ 16 \ 3)(2 \ 12 \ 9 \ 5)(4 \ 14 \ 11 \ 7)(6 \ 15 \ 13 \ 8)$

\noindent $-{\bf T} \mapsto (x^2y,1)\mapsto (-\gamma,1)\mapsto (1735)(2648)\mapsto (1 \ 11 \ 16 \ 4)(2 \ 13 \ 9 \ 6)(5 \ 8 \ 12 \ 15)(10 \ 14 \ 3 \ 7)$

\noindent $-{\bf C*\bf P} \mapsto (x^3,z)\mapsto (-\iota,-1)\mapsto (1432)(5876)(9 \ 10)\mapsto (1 \ 12 \ 16 \ 5)(2 \ 10 \ 9 \ 3)(4 \ 15 \ 11 \ 8)(6 \ 14 \ 13 \ 7)$

\noindent $-{\bf C*\bf T} \mapsto (x^2y,z)\mapsto (-\gamma,-1)\mapsto (1735)(2648)(9 \ 10) \mapsto (1 \ 13 \ 16 \ 6)(2 \ 11 \ 9 \ 4)(5 \ 7 \ 12 \ 14)(8 \ 10 \ 15 \ 3)$

\noindent $-{\bf P*\bf T} \mapsto (x^3y,1)\mapsto (-\kappa,1)\mapsto (1836)(2745)\mapsto (1 \ 14 \ 16 \ 7)(2 \ 15 \ 9 \ 8)(3 \ 11 \ 10 \ 4)(5 \ 13 \ 12 \ 6)$

\noindent $-\Theta \mapsto (x^3y,z)\mapsto (-\kappa,-1)\mapsto (1836)(2745)(9 \ 10)\mapsto (1 \ 15 \ 16 \ 8)(2 \ 14 \ 9 \ 7)(3 \ 13 \ 10 \ 6)(4 \ 5 \ 11 \ 12)$

\noindent $-1\mapsto (x^2,1)\mapsto (-1,1)\mapsto (13)(24)(57)(68)\mapsto (1 \ 16)(2 \ 9)(3 \ 10)(4 \ 11)(5 \ 12)(6 \ 13)(7 \ 14)(8 \ 15)$ $$\eqno{(73)}$$ where $\Z_2\to S^0$ is given by $1 \mapsto 1$ and $z\mapsto -1$, $H\subset S_{10}$, and $K\subset S_{16}$. 

\

As far as $DC_8$ (or $Q$) -and therefore for $G_\Theta$- we can go into a more detailed description by taking into account that $DC_8$ is a {\it hamiltonian} group, that is, all its subgroups are invariant (Herstein, 1996). In particular $\{1,x,x^2,x^3\}\cong \Z_4$ and $\{1,x^2\}\cong \Z_2$, its center, are  proper subgroups. Correspondingly, one has the short exact sequences $$0\to \Z_4 \buildrel {\iota_8}\over \longrightarrow DC_8 \buildrel {\varphi_8}\over \longrightarrow \Z_2 \to 0 \eqno{(74)}$$ and $$0\to \Z_2 \buildrel {\iota_8^\prime}\over \longrightarrow DC_8 \buildrel {\varphi_8^\prime}\over \longrightarrow {{DC_8}\over{\Z_2}}\cong \Z_2\times \Z_2 \to 0 \eqno{(75)}$$ where $\iota_8$ and $\iota_8^\prime$ are the inclusions, $Ker(\varphi_8)=\Z_4$, $\varphi_8(y)=\varphi_8(xy)=\varphi_8(x^2y)=\varphi_8(x^3y)=-1$, $Ker(\varphi_8^\prime)=\{1,x^2\}$, and $\varphi_8^\prime (\alpha)=[\alpha]=\alpha\{1,x^2\}$ with $[1]=\{1,x^2\}, \ [x]=\{x,x^3\}, \ [y]=\{y,x^2y\}$, and $[xy]=\{xy,x^3y\}$. The isomorphism ${{DC_8}\over {\Z_2}}\buildrel {\rho}\over \longrightarrow \Z_2\times \Z_2\equiv V$, the {\it four} or {\it Klein's group}, is given by $\rho([1])=(1,1), \ \rho([x])=(1,-1), \ \rho([y])=(-1,1)$ and $\rho([xy])=(-1,-1)$. Then, $DC_8$ (or $Q$) is an {\it abelian non trivial non central (central) extension of} $\Z_4$  ($\Z_2$) {\it by} $\Z_2$ ($\Z_2\times \Z_2$). None of these extensions however, splits, since, as can be easily verified, it is not possible to define functions $\gamma_8:\Z_2\to DC_8$ and $\gamma_8^\prime:\Z_2\times \Z_2 \to DC_8$, simultaneously being group homomorphisms and satisfying $\varphi_8 \circ \gamma_8=Id_{\Z_2}$ and $\varphi_8^\prime \circ \gamma_8^\prime=Id_{\Z_2\times \Z_2}$. Then, $DC_8$ (and so $Q$) is not a semidirect product, neither of $\Z_4 $ and $\Z_2$ nor of $\Z_2$ and $\Z_2\times \Z_2$ (see e.g. Mac Lane and Birkoff, pp. 414-6). Nevertheless, the geometric content of $Q$ -and therefore of $G_\Theta$- is clear from the geometric content of $\Z_4$, $\Z_2$ and $\Z_2\times \Z_2$.

\

\

{\bf 8. Weyl and Majorana representations}

\

In the Weyl representation of the Dirac algebra, any matrix $A\in D^{16}$ of the standard representation is transformed into $$A_W=S_WAS_W^{\dag}, \eqno{(76)}$$ where $$S_W={{1}\over{\sqrt{2}}}(\gamma_0-\gamma_5)_{DP}={{1}\over{\sqrt{2}}}\pmatrix{1&1 \cr 1&-1}=S_W^{\dag}=S_W^{-1}, \ det(S_W)=-1, \ tr(S_W)=0. \eqno{(77)}$$

In the massless limit, where the left ($\psi_L$) and right ($\psi_R$) parts of the Dirac field defined by $\psi_W=S_W\psi=\pmatrix{\psi_R \cr \psi_L}$ decouple from each other and obey $({{\partial}\over {\partial t}}\mp\vec{\sigma}\cdot \nabla)\psi_{L,R}=0$. The discrete transformations are given by $$C_W^{(1)}=\pm \pmatrix{\sigma_2&0 \cr 0&-\sigma_2 \cr}, \ P_W^{(1)}=\pm i\pmatrix{0&1 \cr 1&0 \cr}, \ T_W^{(1)}=\pm \pmatrix{\sigma_2&0 \cr 0&\sigma_2}, \ \theta_W^{(1)}=\pm i\pmatrix{0&1 \cr -1&0 \cr} \eqno{(78)}$$ in $G_\theta^{(1)}$, and $$C_W^{(2)}=iC_W^{(1)}, \ P_W^{(2)}=P_W^{(1)}, \ T_W^{(2)}=iT_W^{(1)}, \ \theta_W^{(2)}=-\theta_W^{(1)} \eqno{(79)}$$ in $G_\theta^{(2)}$.

\

It is clear that the group structures of subsection 7.1 are preserved, and that the derivations in sections 2 to 6 are independent of the mass. For example, $T_WC_W=S_WTS^{-1}_WS_WCS_W^{-1}=S_WTCS_W^{-1}=S_WCTS_W^{-1}=C_WT_W.$

\

In the Majorana representation, any matrix $A\in D^{16}$ in the standard representation is transformed into $$A_M=S_MAS_M^{\dag} \eqno{(76a)}$$ where $$S_M={{1}\over {\sqrt{2}}}(\gamma^2\gamma_0+\gamma_0)_{DP}={{1}\over{\sqrt{2}}}\pmatrix{1&-\sigma_2 \cr -\sigma_2 &-1 \cr}=S^{\dag}_M=S^{-1}_M, \ det(S_M)=1, \ tr(S_M)=0. \eqno{(77a)}$$ This transformation is such that all gamma matrices become pure imaginary: $$\gamma_{0M}=\pmatrix{0&0&0&i\cr 0&0&-i&0\cr 0&i&0&0&\cr -i&0&0&0 \cr}, \ \gamma^1_M=\pmatrix{-i&0&0&0\cr 0&i&0&0\cr 0&0&-i&0\cr 0&0&0&i\cr},$$ $$\gamma^2_M=\pmatrix{0&0&0&i\cr 0&0&-i&0\cr 0&-i&0&0\cr i&0&0&0\cr}, \ \gamma^3_M=\pmatrix{0&i&0&0\cr i&0&0&0\cr 0&0&0&i\cr 0&0&i&0\cr}. \eqno{(80)}$$ For the discrete transformations one obtains: $$C_M^{(1)}=\pm \pmatrix{1&0\cr 0&-1}, \ P_M^{(1)}=\pm i\pmatrix{0&-\sigma_2 \cr -\sigma_2 &0\cr}, \ T_M^{(1)}=\pm \pmatrix{\sigma_2 &0\cr 0&\sigma_2\cr}, \ \theta_M^{(1)}=\pm i\pmatrix{0&-1\cr 1&0\cr}\eqno{(78a)}$$ in $G_\theta^{(1)}$, and $$C_M^{(2)}=iC_M^{(1)}, \ P_M^{(2)}=P_M^{(1)}, \ T_M^{(2)}=iT_M^{(1)}, \ \theta_M^{(2)}=-\theta_M^{(1)} \eqno{(79a)}$$ in $G_\theta^{(2)}$.

\

As in the previous case, the group structures of subsection 7.1 are preserved. For example, $(CP)_MP_M$

$=S_MCPS_M^{-1}S_MPS_M^{-1}=S_MCP^2S_M^{-1}=-S_MCS_M^{-1}=-C_M.$ 

\

{\bf Acknowledgements}

\

The author thanks the hospitality of the Instituto de Astronom\'\i a y F\'\i sica del Espacio (IAFE), UBA-CONICET, Buenos Aires, Argentina, and of the Departamento de F\'\i sica Te\'orica de la Facultad de Ciencias F\'\i sicas de la Universidad de Valencia (UV), Espa$\tilde{n}$a, where part of this work was performed. Also, he thanks professors Rafael Ferraro (IAFE) and Jos\'e A. de Azc\'arraga (UV) for useful discussions, and the graduate student Eric Mart\'\i nez (UNAM) for a valuable comment. 

\

\

{\bf References}

\

Aguilar, M. A. and Socolovsky, M. (1997). Naturalness of the Space of States in Quantum Mechanics, {\it International Journal of Theoretical Physics} {\bf 36}, 883-921.

\

Armstrong, M. A. (1988). {\it Groups and Symmetry}, Springer-Verlag, New York: p. 71.

\

Asche, D. (1989). {\it An Introduction to Groups}, Adam Hilger, Bristol: p. 86. 

\

de Azc\'arraga, J. A. and Boya, L. J. (1975). On the Particle-Antiparticle Conjugation, {\it Reports on Mathematical Physics} {\bf 7}, 1-8.

\

de Azc\'arraga, J. A. (1975). {\it P, C, T, $\theta$ in Quantum Field Theory}, GIFT 7/75, Zaragoza, Spain.

\

Berestetskii, V. B., Lifshitz, E. M. and Pitaevskii, L. P. (1982). {\it Quantum Electrodynamics, Landau and Lifshitz Course of Theoretical Physics, Vol. 4}, 2nd. edition, Pergamon Press, Oxford: pp. 69-70.

\

Bogoliubov, N. N. and Shirkov, D. V. (1980). {\it Introduction to the Theory of Quantized Fields}, 3rd. edition, Wiley, New York: pp. 87-8.

\

Capri, A. Z. (2002). {\it Relativistic Quantum Mechanics and Introduction to Quantum Field Theory}, World Scientific, New Jersey: pp. 46-51.

\

Feynman, R. P. (1987). The Reason for Antiparticles, in {\it Elementary Particles and the Laws of Physics. 1986 Dirac Memorial Lectures}, eds. R. P. Feynman and S. Weinberg, Cambridge University Press, New York: p. 41.

\

Hamermesh, M. (1989). {\it Group Theory and its Application to Physical Problems}, Dover, N. Y.: p. 16.

\

Herstein, I. N. (1996). {\it Abstract Algebra}, 3rd. ed., Prentice Hall, New Jersey: p. 72.

\

Lawson, H. B. and Michelsohn, M. L. (1989). {\it Spin Geometry}, Princeton University Press, Princeton, New Jersey: p. 14.

\

Mac Lane, S. and Birkoff, G. (1979). {\it Algebra}, 2nd. ed., Macmillan Pub. Co., New York: p. 413.

\

Porteous, I. R. (1981). {\it Topological Geometry}, Cambridge University Press, Cambridge: p. 240.

\

Racah, G. (1937). Sulla Simetria tra Particelle e Antiparticelle, {\it Nuovo Cimento} {\bf 14}, 322-328.

\

Sakurai, J. J. (1985). {\it Modern Quantum Mechanics}, Benjamin, Menlo Park, California: p. 278.

\

S\'anchez Valenzuela, O. A. (1991). Matem\'atica de las Simetr\'\i as Discretas de la F\'\i sica, {\it Ciencia} {\bf 42}, 125-140.

\

Schweber, S. S. (1961). {\it An Introduction to Relativistic Quantum Field Theory}, Harper and Row, New York: pp. 105-7.

\

Shirokov, Y. M. (1958). A Group-Theoretical Consideration of the Basis of Relativistic Quantum Mechanics. IV. Space Reflections in Quantum Theory, {\it Soviet Physics JETP} {\bf 34}, 493-498.

\

Shirokov, Y. M. (1960). Space and Time Reflections in Relativistic Theory, {\it Nuclear Physics} {\bf 15}, 1-12.

\

Socolovsky, M. (2001). On the Geometry of Spin ${{1}\over {2}}$, {\it Advances in Applied Clifford Algebras} {\bf 11}, 487-494.

\

Sternberg, S. (1987). On Charge Conjugation, {\it Communications in Mathematical Physics} {\bf 109}, 649-679. 

\

Sternberg, S. (1994). {\it Group Theory and Physics}, Cambridge University Press, Cambridge: p. 152.

\

Varlamov, V. V. (2003). Group Theoretical Interpretation of CPT-theorem, arXiv: math-ph/0306034.

\

Wick, G. C. and Wightman, A. S. and Wigner, E. P. (1952). The Intrinsic Parity of Elementary Particles, {\it Physical Review} {\bf 88}, 101-105. 

\

Yang, C. N. and Tiomno, J. (1950). Reflection Properties of Spin ${{1}\over {2}}$ Fields and a Universal Fermi-Type Interaction, {\it Physical Review} {\bf 79}, 495-498.

\

{\bf Appendix 1}

\

{\it Derivation of} (7):

\

Writing $P=\pmatrix{A&B\cr C&D\cr}$ with $A,B,C,D\in \C(2)$, $P\gamma_0=\gamma_0P$ implies $B=C=0$ {\it i.e.} $P=\pmatrix{A&0 \cr 0&D \cr}$. Then $P\gamma^1=-\gamma^1 P$ implies $D=-\sigma_1A\sigma_1$ and so $P=\pmatrix{A&0 \cr 0& -\sigma_1A\sigma_1 \cr}$. Writing $A=\pmatrix{a&b\cr c&d \cr}$ with $a,b,c,d\in \C$, $P\gamma^2=-\gamma^2P$ implies $b=c=0$ and so $A=\pmatrix{a&0 \cr 0&d\cr}$ and $D=\pmatrix{-d&0\cr 0&-a \cr}$. Finally, $P\gamma^3=-\gamma^3P$ implies $d=a$ and so $P=a\gamma_0$ with $a\in\C^*$. 

\

{\it Derivation of} (18):

\

Writing $C=\pmatrix{A&B \cr E&F \cr}$ with $A,B,E,F\in \C(2)$, $C\gamma_0=-\gamma_0C$ implies $A=F=0$ and therefore $C=\pmatrix{0&B \cr E&0 \cr}$. Then $C\gamma^2=-\gamma^2C$ implies $E=\sigma_2B\sigma_2$ and so $C=\pmatrix{0&B \cr \sigma_2B\sigma_2 &0\cr}$. Then $C\gamma^1=\gamma^1C$ implies $\sigma_1B\sigma_1=-\sigma_2B\sigma_2$ (*) and $C\gamma^3=\gamma^3C$ implies $\sigma_2B\sigma_2=-\sigma_3B\sigma_3$ (**). Writing $B=\pmatrix{\alpha&\beta \cr \gamma&\delta \cr}$ with $\alpha,\beta,\gamma,\delta \in \C$, (*) and (**) imply $\alpha=\delta=0$ and $\gamma=-\beta$. Then $B=\beta\pmatrix{0&1 \cr -1&0\cr}=i\beta\sigma_2$ and so $C=i\beta\pmatrix{0&\sigma_2 \cr \sigma_2&0 \cr}=\eta\gamma^2\gamma_0$ with $\eta=-i\beta\in \C^*$.

\

{\it Derivation of} (27):

\

Writing $T=\pmatrix{A&B\cr C&D\cr}$ with $A,B,C,D\in \C(2)$, $T\gamma_0=\gamma_0T$ implies $B=C=0$ and so $T=\pmatrix{A&0\cr 0&D\cr}$. Then $T\gamma^1=-\gamma^1T$ (since $\gamma^{1*}=\gamma^1$) implies $D=-\sigma_1A\sigma_1$ and so $T=\pmatrix{A&0\cr 0&-\sigma_1A\sigma_1 \cr}$, and $T\gamma^3=-\gamma^3T$ (since $\gamma^{3*}=\gamma^3$) implies $A=\sigma_2A\sigma_2$ (*). Writing $A=\pmatrix{a&b\cr c&d \cr}$, (*) implies $c=-b$ and $d=a$ {\it i.e.} $A=\pmatrix{a&b\cr -b&a\cr}$. Finally, since $\gamma^{2*}=-\gamma^2$, $T\gamma^2=\gamma^2T$ leads to $A=-\sigma_3A\sigma_3$ which implies $a=0$ {\it i.e.} $A=b\pmatrix{0&1\cr -1&0\cr}=ib\sigma_2$ and so $T=z\pmatrix{\sigma_2 & 0 \cr 0&\sigma_2 \cr}=w\gamma^3\gamma^1$ with $w=iz\in \C^*$.

\

{\bf Appendix 2}

\

{\it Derivation of} (68) and (69):

\

Let $\psi_i(t,\vec{x})$ denote the i-th component of the Dirac wave function $\psi(t,\vec{x})$, $i=1,2,3,4$; then:

\

i) $\psi_{iC\Pi}(t,\vec{x})=(\psi_{iC})_\Pi(t,\vec{x})=P_{ij}\psi_{jC}(t,-\vec{x})=P_{ij}(C\gamma_0)_{jk}\psi_k(t,\vec{x})^*=(PC\gamma_0)_{ik}\psi_k(t,\vec{x})^*$,

$\psi_{i\Pi C}(t,\vec{x})=(\psi_{i\Pi})_C(t,\vec{x})=(C\gamma_0)_{ij}\psi_{j\Pi}(t,\vec{x})^*=(C\gamma_0)_{ij}P_{jk}^*\psi_k(t,-\vec{x})^*=-(C\gamma_0)_{ij}P_{jk}\psi_k(t,-\vec{x})^*=-(C\gamma_0P)_{ik}\psi_k(t,-\vec{x})^*=-(CP\gamma_0)_{ik}\psi_k(t,-\vec{x})^*=(PC\gamma_0)_{ik}\psi_k(t,-\vec{x})^*$; {\it i.e.} $\psi_{iC\Pi}(t,\vec{x})=\psi_{i\Pi C}(t,\vec{x})$. 

On the other hand, for the corresponding field operators, one has:

$\psi_{iC\Pi}(t,\vec{x})={\bf P} ^{\dag}\psi_{iC}(t,\vec{x}){\bf P}={\bf P}^{\dag}{\bf C}^{\dag}\psi_i(t, \vec{x}){\bf C \bf P}=({\bf C \bf P})^{\dag}\psi_i(t,\vec{x})({\bf C \bf P})=({\bf C*\bf P})\cdot \psi_i(t, \vec{x})$,

$\psi_{i\Pi C}(t,\vec{x})={\bf C}^{\dag}\psi_{i\Pi}(t,\vec{x}){\bf C}={\bf C}^{\dag}{\bf P}^{\dag}\psi_i(t,\vec{x}){\bf P \bf C}=({\bf P \bf C})^{\dag}\psi_i(t,\vec{x})({\bf P \bf C})=({\bf P*\bf C})\cdot \psi_i(t,\vec{x})$. 

Then, consistency between the one particle theory and the quantum field theory, implies $({\bf C*\bf P})\cdot \psi_i(t,\vec{x})=({\bf P*\bf C})\cdot \psi_i(t,\vec{x})$ and from the arbitrariness of $\psi_i(t,\vec{x})$, ${\bf C*\bf P}={\bf P*\bf C}$. 

\

ii) $\psi_{iC\tau}(t,\vec{x})=(\psi_{iC})_\tau(t,\vec{x})=T_{ij}\psi_{jC}(-t,\vec{x})^*=T_{ij}(C\gamma_0)^*_{jk}\psi_k(-t,\vec{x})=T_{ij}(C^*\gamma_0)_{jk}\psi_k(-t,\vec{x})$

$=(TC^*\gamma_0)_{ik}\psi_k(-t,\vec{x})=\mp(TC\gamma_0)_{ik}\psi_k(-t,\vec{x})$, where the - and + signs respectively refer to the solutions (21a) and (21b) for $C$;

$\psi_{i\tau C}(t,\vec{x})=(\psi_{i\tau})_C(t,\vec{x})=(C\gamma_0)_{ij}\psi_{j\tau}(t,\vec{x})^*=(C\gamma_0)_{ij}T_{jk}^*\psi_k(-t,\vec{x})=(C\gamma_0T^*)_{ik}\psi_k(-t,\vec{x})$ 

$=\mp(C\gamma_0T)_{ik}\psi_k(-t,\vec{x})=\mp(CT\gamma_0)_{ik}\psi_k(-t,\vec{x})=\mp(TC\gamma_0)_{ik}\psi_k(-t,\vec{x})$, where the - and + signs respectively refer to the solutions (32) and (33) for $T$; then $\psi_{iC\tau}(t,\vec{x})=\psi_{i\tau C}(t,\vec{x})$. 

On the other hand, for the quantum field operators, one has:

$\psi_{C\tau}(t,\vec{x})={\bf T}^{\dag}\psi_C(t,\vec{x}){\bf T}={\bf T}^{\dag}({\bf C}^{\dag}\psi(t,\vec{x}){\bf C}) {\bf T}=({\bf C \bf T})^{\dag}\psi(t,\vec{x})({\bf C \bf T})=({\bf C}*{\bf T})\cdot \psi(t,\vec{x})={\bf T}^{\dag}C\bar{\psi}(t,\vec{x})^{\sim}{\bf T}$ 
$=C^*{\bf T}^{\dag}\bar{\psi}(t,\vec{x})^{\sim}{\bf T}=C^*T\bar{\psi}(-t,\vec{x})^{\sim};$

$\psi_{\tau C}(t,\vec{x})={\bf C}^{\dag}\psi_{\tau}(t,\vec{x}){\bf C}={\bf C}^{\dag}({\bf T}^{\dag}\psi(t,\vec{x}){\bf T}) {\bf C}=({\bf T \bf C})^{\dag}\psi(t,\vec{x})({\bf T \bf C})=({\bf T}*{\bf C})\cdot \psi(t,\vec{x})={\bf C}^{\dag}T\psi(-t,\vec{x}){\bf C}$ $=T{\bf C}^{\dag}\psi(-t,\vec{x}){\bf C}=TC\bar{\psi}(-t,\vec{x})^{\sim}.$

Consistency with the one particle theory implies ${\bf C}^{\dag}({\bf T}^{\dag}\psi(t,\vec{x}){\bf T}){\bf C}={\bf T}^{\dag}({\bf C}^{\dag}\psi(t,\vec{x}){\bf C}){\bf T}$ {\it i.e.} ${\bf T}*{\bf C}={\bf C}*{\bf T}$ and then $TC=C^*T$, that is, $CT^*=T^*C^*$. From (31), $T^*C^*=TC^*$ and therefore  $T^*=T$. 

\

\

\

\

\

\

\

\

\

e-mail: socolovs@nuclecu.unam.mx

\end